# Reflections on the search for particle dark matter by direct experiments

Alessandro Bottino[1]
*University of Torino, Italy*
*Accademia delle Scienze di Torino, Italy*

## Abstract

*Since the daring intuition by Fritz Zwicky in 1933 about the existence of dark matter in the Universe, in spite of the extensive investigations pursued over a very long lapse of time, the nature of this matter and of its (other than gravitational, if any) interactions have remained unknown. Very likely, at least a fraction of this matter consists of fossil particles; a possibility that in the last decades has prompted direct and indirect searches for these relics. Particular attention has been (and is being) devoted to the investigation about the physical effects that fossil particles, and in particular Weakly Interacting Particles (WIMPs), moving in our galaxy can produce when they scatter off the material of an appropriate underground detector. In this note we consider the present status of this type of experimental investigation, and comment about some important results obtained recently by the DAMA Collaboration and their possible developments.*

## 1. Introduction

An impressive host of astronomical and astrophysical observations (flatness of the rotation curves of single galaxies, properties of galaxy clusters, features of the cosmic microwave background, and others), accumulated along more than eighty years, supports the brilliant and daring intuition by Fritz Zwicky in 1933 [1] about the existence of a large amount of dark matter (DM) in the Universe. The theory of formation of cosmological structures provides further compelling support to the Zwicky's conjecture.

The presence in the Cosmos of a number of relics from the Big Bang, beyond ordinary matter (protons, neutrons, electrons) and neutrinos, looks quite natural. Indeed, many theories of particle physics and descriptions of cosmic evolution imply that, by various mechanisms, some kinds of particles (such as axions, neutralinos, sterile neutrinos) are present as fossils in our Universe, together with cosmological objects (typically, primordial black holes). Thus each of these physical entities, if really existing, can legitimately be considered as DM.

The most remarkable aspect is that DM, on the grounds of the mentioned observations, turns out to be even more abundant than ordinary matter, roughly by a factor of about 5 [2]. The total DM contribution might be due prominently to one single relic object, with a number of other fossils contributing very little to the total cosmological amount of DM. Though, it is obviously not excluded that more than one component share the role of sizeable DM components; a situation which would however raise the question whether this circumstance would be accidental or a sign of some correlation among different components.

Independently of the actual contribution to the total amount of DM, the direct search for each fossil candidate is obviously of great importance for our comprehension of cosmology, astrophysics, and particle physics.

---
[1] email: aless.bottino@gmail.com; http//:www.alessandrobottino.it

In the present note only particle fossils are considered, and more specifically those which are generically denominated as WIMP (Weakly Interacting Massive Particles), even though we do not imply for them a dominant role as components of DM. We do not enter into specific physical realizations of WIMPs, except for some occasional mention of one of the favorite candidates: the neutralino [3].

This note is not meant to be a thorough overview of the field. Rather, it puts the emphasis on some aspects of the field that appear to be of particular relevance, but are often overlooked in current literature.

Our comments on the very last developments are preceded by a retrospective look at the first steps in the field. In Sect.2 we outline some general features of the strategies employed in direct search of WIMPs and the developments focused on measurements of the expected WIMP annual modulation. In Sect.3 we comment on other experimental avenues. Sect. 4 is devoted to a discussion on the main results obtained recently by measurements of the annual modulation effects with low energy threshold. In Sect. 5 we present some final remarks.

## 2. Direct search for WIMPs: beginnings and developments

Experimental searches for the existence of fossil particles in our galaxy are best performed by analyzing the effects induced by these relics when they scatter off the material contained in an appropriate detector placed in a low-background environment. This line of experimental activity was prompted in the second half of the Eighties by a series of theoretical seminal papers.

### 2.1 Principle of detection and signal signatures

In 1985 Goodman and Witten [4] considered the possibility that the kind of detectors, originally suggested by Drukier and Stodolsky to measure solar and reactor neutrinos [5], could be employed to measure certain classes of relic particles. In particular they considered the case of a particle interacting coherently with the atomic nucleus (a possible example within a supersymmetric theory being the sneutrino) and the case of a particle interacting with a nucleus with a spin-dependent coupling (a possible example within a supersymmetric theory being the photino). Actual estimates for the relevant detection rates were performed in Ref. [4].

This kind of evaluations was further developed by Drukier, Freese and Spergel [6], who also noticed that *the Earth's motion around the Sun would produce a distinctive modulation in the signal detected from halo particle candidates*. This remarkable feature was emphasized as capable of playing a crucial role in discriminating signal from backgrounds. Further investigations of the signal modulation were presented in Refs. [7-8].

These last theoretical considerations were fundamental in putting the emphasis on the fact that the most natural and efficient approach in the search for DM particles in our galaxy consists in looking for effects due to the relative velocity of the Earth with the DM particle wind, *i.e.* annual and diurnal variations of the signal and its directionality. Use of these features provides irreplaceable tools in the experimental search for two reasons: 1) it represents a unique way to prove that a measured signal is genuine (provided that no other similarly-modulated background is present in the experimental environment), 2) it strongly suppresses the most common backgrounds. Whereas signal directionality is under study, but not yet measurable in present set-ups, signal annual modulation has already been systematically investigated, as discussed here below. Improved sensitivities of present devices and/or other experimental set-ups are likely to provide information about diurnal modulation in a near future.

### 2.2 Getting ready for the exploration of a popular DM candidate: the neutralino

Some overviews of the experimental beginnings of direct detection towards the end of the Eighties and the beginning of the Nineties are described for instance in Ref.[9]. Typically, classical Ge (Si) semiconductor detectors were able to investigate some of particle candidates with oddly high interaction rates with ordinary matter (cosmions, heavy Dirac neutrinos). However, the race for a large sample of various new set-ups (crystal and liquid scintillators, bolometers, track projection chambers, and others) had started.

An important experimental breakthrough occurred when it was shown that an efficient DM particles search by using a set-up made of ultra-low-background NaI(Tl) crystals deep underground was feasible [10-11]. It was however important to estimate which should be the size (together with other relevant features) of the NaI set-up for its sensitivity to be commensurate with the expected rates for the detection on some *realistic* theoretical model. Since great expectations focused on candidates provided by supersymmetric theories, in Ref.[12] a detailed comparison was carried out to establish how current experimental sensitivity of the NaI set-up compared with the expected neutralino detection rates. Soon it turned out that the experimental target for a detector based on NaI crystals would be to reach a size of order of 100 kg. Such a set-up was actually developed and installed in the Gran Sasso Laboratory [13]; after an exposure of 4123 Kg day [14], it reached a sensitivity adequate to perform an experimental investigation of a sizeable portion of the neutralino physical (*i.e.*, not yet excluded by accelerator searches) parameter space, as remarked in Ref.[15].

### 2.3 Towards the measurement of the WIMP annual modulation

At the TAUP97 conference a first indication of an annual modulation effect in the search for WIMPs was announced by the DAMA Collaboration [16-17]. Compatibility of these preliminary data with possible signals due a neutralino candidate was reported at the same Conference [18-19].

Subsequent phenomenological analyses, extended to WIMP masses smaller than the usually considered lower bound for the neutralino mass, led to the conclusion that a NaI detector of the capability of the DAMA set-up would have a marked sensitivity in two specific ranges for the WIMP mass: one around 10 GeV and one around 50-100 GeV [20]. It is worth noting that the actual values of the two characteristic mass intervals depend sensitively on the WIMP galactic distribution function (DF) and on parameters describing the detector nuclear properties. This is illustrated in Fig.4 of Ref. [20], where a few examples of DFs are considered, among those discussed in detail in Ref.[21]. These features, involving light WIMPs (i.e., WIMPs with a mass between a few GeV and a few tens of GeV) became a commonly discussed property in the context of subsequent phenomenological analyses performed by various authors on NaI-detector results.

Since the time of the report of a first indication of annual modulation, the DAMA Collaboration progressively increased the statistical significance of this effect through successive improvements of the original set-up and an impressive collection of data. In 2003 the evidence of the annual variation reached the statistical level of 6.3 σ C.L. [22-23]. These data, when analyzed in terms of a WIMP with a coherent scattering, singled out (in a plot of the WIMP-nucleon cross section vs WIMP mass) a region reminiscent of the sensitivity property anticipated in Ref.[20], as pointed out in Ref. [24]. In this last paper it was proved that light WIMPs, under the features of neutralinos, compared quite well with the DAMA annual modulation results.

Through the sizeable increase in the detector mass (from a mass ≈ 100 Kg of DAMA/NaI to a mass ≈ 250 Kg of DAMA/LIBRA - phase 1), various improvements in the experimental apparatus and an impressive temporal exposure (14 annual cycles), the DAMA Collaboration was able to reach a total exposure of 1.33 ton year and a statistical significance for the annual modulation effect of 9.3 σ [25-26]. It is important to stress that these results were made possible by the remarkable stability of the operational experimental parameters of the DAMA set-ups.

All along the evolution of the DAMA annual modulation measurements these results have prompted a lot of theoretical and phenomenological analyses covering a rather large range of different scenarios. Among others: light WIMPs (including neutralinos) [27-34], mirror dark matter [35-36], inelastic dark matter [37-39]. Various phenomenological scenarios were investigated by the DAMA Collaboration (see, for instance, Ref.[40]).

Further results have been obtained recently by the DAMA Collaboration with an improved configuration, DAMA/LIBRA - phase 2, which allows a lower software energy threshold of 1 KeV as compared to the previous one of 2 KeV [41-42]. These new data, collected over 6 annual cycles, added to those of DAMA/NaI and DAMA/LIBRA – phase 1, provide a statistical significance of the annual modulation of 12.9 σ. We postpone the discussion of these results to Sect. 4.

Other experiments of direct detection operating with NaI crystals are those of Refs. [43-47]. The first three of these experiments plan to get an exposure adequate for measuring the annual modulation.

### 2.3.1 Can backgrounds of various origins mimic the DM annual modulation effect?

In view of the high level of statistical significance of the DAMA results, the only point that could invalidate the interpretation in terms of a genuine signal of a DM particle is the possibility that some background might vary with all the same peculiarities as the signature.

The very strict provisos set by the DAMA Collaboration for the definition of the genuine signal, set already from the very beginning, put quite severe constraints on possible backgrounds. However, because of the high relevance of this point, a deep inspection of alternative (not DM-related) explanations of the annual modulation, as also suggested by a large number of authors, has always been a matter of accurate estimates performed by the DAMA Collaboration for any imaginable peculiar modulating background, with the conclusion that all these possibilities would produce effects well below the signal level and unable to satisfy all the peculiarities of the exploited signature. The same conclusion has been reached by DAMA Collaboration by examining other suggested origin of spurious effects. Detailed discussions on this matter and related references can be found *e. g.* in Refs.[48-49] and references quoted therein.

### 3. Signal/background discrimination by measurements of two signals

At variance with the NaI detectors mentioned so far, other experiments do not aim at cutting down the background by measuring the annual modulation expected for the signal, but try to discriminate DM signal from background by measuring two signals of different nature: typically, scintillation and phonons/heat, or scintillation and ionization. Experiments of this category are, among others, CRESST-III [50], XENON 1T [51], LUX [52].

An important drawback of this detection strategy is that one renounces to make use of a specific signature of the searched-for effect, with the dramatic consequence that large backgrounds have to be estimated and subtracted from the collected data, using simulation-based background modelling.

This implies that the very constraining upper bounds to the WIMP-ordinary matter cross sections reported by the previous experiments should actually be taken with great caution.

The possible flaws intrinsic in the adopted experimental strategy appear to be often underestimated in current literature, with the consequence that null results are frequently accepted without deeper scrutiny. It is also worth recalling that comparing results of experimental set-ups using different target materials is uncertain because of the different nuclear properties.

### 4. DAMA/LIBRA phase 2 data

As mentioned above the DAMA Collaboration has recently published the results of the new experimental configuration DAMA/LIBRA - phase 2, in which the software energy threshold has been lowered to 1 KeV from the previous one of 2 KeV [42]. These results, anticipated in [41], have immediately attracted much interest, because of the relevant information provided by the features of the energy spectrum in the energy range (1-2) KeV.

In Refs.[53-54] the new DAMA energy spectrum is analyzed in terms of WIMPs with both spin-independent (SI) and spin-dependent (SD) interactions with the nucleons, assuming a WIMP Gaussian isotropic distribution in the galactic rest frame. Both canonical (isospin conserving) and isospin-violating interactions are considered. In Ref.[54] the analysis is further extended to WIMP-nucleus interactions depending on the WIMP velocity, which arise in a general non-relativistic Effective Field Theory [55-56].

The analyses of Refs.[53-54], that go in particular through a rebinning of the original experimental data, bring to similar conclusions, i.e. that, in the case of a SI interaction of low-mass candidates a good fit is obtained only by allowing a sizeable isospin-violation effect. Good fits are instead found in the case of a spin-spin interaction, also with isospin conservation. Some other interaction terms considered in Ref.[54] have similar good fits without requiring isospin violation.

Other phenomenological interpretations of the DAMA/LIBRA – phase 2 data have been suggested by other authors, for instance, in Ref.[57] and Ref.[58].

A systematic analysis of how DM models compare with the DAMA/LIBRA-phase2 results has been performed by the DAMA Collaboration [59-61]. This analysis covers the two possible interaction mechanisms: one due to the interaction of the DM particle with atomic electrons, the other due to the interaction of the DM particle with the target nuclei. For these last instance various options for the quenching factors are examined. Furthermore, a large variety of particle DM candidates are worked out, employing also a selection of different analytic forms for the WIMP galactic distribution function. It is also worth stressing that, at variance with other analyses, the employed $\chi^2$ includes a term encoding the experimental bounds on the unmodulated part of the signal. Also the channeling effect in the crystals is taken into account.

In Refs.[59-60] it is found that many different models can provide a good fit to the DAMA/LIBRA-phase2 results (see, in particular, slide 24 of Ref.[59] for a few examples), the two most remarkable features being that: a) in virtue of the lower energy threshold, for each model the new data restrict significantly the model parameter space, b) even the simplest *classical* case of a coherent WIMP-nucleus interaction with a light WIMP mass fits remarkably well the experimental data (even for an classical isothermal galactic distribution, without any need for isospin-violation). This last result is at variance with what found in Refs.[53-54]; a fact most likely due to important differences in the features of the different analyses (quenching factors, binning, $\chi^2$ definition, and other).

## 5. Final remarks

As we have seen in the previous sections, the measurement of the annual modulation effect performed by the DAMA Collaboration provides an impressive indication that at least a fraction of the DM in our Universe may consist of relic particles which interact with ordinary matter by some weak force. The mass of these WIMPs might preferentially fall on the lower side of the mass range (10-100) GeV (somewhat higher values for some distribution functions).

Though other scenarios are possible, it is remarkable that the two most natural cases, originally considered in the seminal paper by Goodman and Witten, of a WIMP with a (isospin conserving) coherent interaction with nuclei or of a WIMP with a spin dependent interaction perfectly fit the annual modulation data. The features of the measured annual variation agree well with the prediction by Drukier, Freese, and Spergel, either using the canonical isotropic galactic distribution or other alternatives.

For the time being no other specific feature of the WIMP, beyond those already mentioned, can be sorted out, and many different scenarios are still possible for different DM candidates. However, it is encouraging that the lowering of the energy threshold turns out to be very effective in restricting the parameter space in each specific scenario. This means that the strategy of improving the investigation of modulation effects in direct detection is a winning one and should be further intensively pursued, with the twofold purpose of further decreasing the energy threshold and of pursuing the investigation of a diurnal variation.

## Acknowledgements

The author is grateful to Rita Bernabei and Pierluigi Belli for information about DAMA results and for useful discussions.

## References

[1] F. Zwicky, Helv. Phys. Acta **6** (1933) 110

[2] For a comprehensive overview on DM see, for instance, K. Freese: *Status of Dark Matter in the Universe,* arXiv:1701.01840 [astro-ph.CO]

[3] Possible DM candidates within supersymmetric theories are thoroughly discussed in M. Drees, R.M. Godbole and P. Roy: *Theory and Phenomenology of Sparticles*, World Scientific, 2004

[4] M.W. Goodman and E. Witten, Phys. Rev. D **31** (1985) 3059

[5] A. Drukier and L. Stodolsky, Phys. Rev. D **30** (1985) 2295

[6] A.K. Drukier, K. Freese, and D.N. Spergel, Phys. Rev. D **33** (1986) 3495

[7] D. N. Spergel, Phys. Rev. D **37** (1988) 1353

[8] K. Freese, J. Frieman and A. Gould, Phys. Rev. D **37** (1988) 3388

[9] Talks given in the DM Section in Proceedings of TAUP91 (Editors A. Morales, J. Morales and J.A. Villar), Nuclear Physics B (Proc. Suppl.) **28**A (1992)

[10] C. Bacci *et al.*, Phys. Lett. B **293** (1992) 460

[11] P. Belli, R. Bernabei, C. Bacci, A. Incicchitti, R. Marcovaldi, and D. Prosperi, DAMA Proposal to INFN Scientific Committee, April 24, 1990

[12] A. Bottino *et al.*, Phys. Lett. B **295** (1992) 330

[13] P. Belli, Proceedings of TAUP95 (Editors A. Morales, J. Morales and J.A. Villar), Nuclear Physics B (Proc. Suppl.) **48** (1996) 60

[14] R. Bernabei *et al.*, Phys. Lett. B **389** (1996) 757

[15] A. Bottino, F. Donato, G. Mignola, S. Scopel, P. Belli, and A. Incicchitti, Phys. Lett. B **402** (1997) 113 [arXiv:hep-ph/9612451]

[16] R. Bernabei *et al.*, Proceedings of TAUP97 (Editors A. Bottino, A. Di Credico and P.Monacelli), Nuclear Physics B (Proc. Suppl.) **70** (1999) 79

[17] R. Bernabei et al., Phys. Lett. B **424** (1998) 195

[18] F. Donato, Proceedings of TAUP97 (Editors A. Bottino, A. Di Credico and P.Monacelli), Nuclear Physics B (Proc. Suppl.) **70** (1999) 117

[19] A. Bottino, F. Donato, N. Fornengo, and S. Scopel, Phys. Lett. B **423** (1998) 109 [arXiv:hep-ph/9709292]

[20] A. Bottino, F. Donato, N. Fornengo, and S. Scopel, Phys. Rev. D **68** (2003) 043506 [arXiv:hep-ph/0304080]

[21] P. Belli, R. Cerulli, N. Fornengo, and S. Scopel, Phys. Rev. D **66** (2002) 043503 [arXiv:hep-ph/0203242]

[22] R. Bernabei *et al.*, Rivista Nuovo Cimento **26** (1) (2003) 1

[23] R. Bernabei *et al.*, Int. J. Mod. Phys. D **13** (2004) 2127

[24] A. Bottino, F. Donato, N. Fornengo, and S. Scopel, Phys. Rev. D **69** (2004) 037302 [arXiv:hep-ph/0307303]

[25] R. Bernabei *et al.*, Eur. Phys. J. C **73** (2013) 2648 [arXiv:1308.5109 [astro-ph.GA]]

[26] R. Bernabei *et al.*, Int. J. Mod. Phys. A **31** no.31 (2016) 1642006

[27] P. Gondolo and G. Gelmini, Phys. Rev. D **71** (2005) 123520 [arXiv:hep-ph/0504010]

[28] F. Petriello and K.M. Zurek, JHEP **09** (2008) 047 [arXiv:0806.3989 [hep-ph]]

[29] A. Bottino, F. Donato, N. Fornengo, and S. Scopel, Phys. Rev. D **78** (2008) 083520 [arXiv:0806.4099 [hep-ph]]

[30] S. Chang, A. Pierce and N. Weiner, Phys. Rev. D **79** (2009) 115011 [arXiv:0808.0196 [hep-ph]]

[31] M. Fairbairn and T. Schwetz, JCAP **0901** (2009) 037 [arXiv:0808.0704 [hep-ph]]

[32] C. Savage, G. Gelmini, P. Gondolo, and K. Freese, JCAP **0904** (2009) 010 [arXiv:0808.3607 [astro-ph]]

[33] A. Bottino, F. Donato, N. Fornengo, and S. Scopel, Phys. Rev. D **81** (2010) 107302 [arXiv:0912.4025 [hep-ph]]

[34] N. Fornengo, S. Scopel, and A. Bottino, Phys. Rev. D **83** (2011) 015001 [arXiv:1011.4743 [hep-ph]]

[35] R. Foot, Phys. Lett. B **728** (2014) 45  [arXiv:1305.4316 [astro-ph.CO]]

[36] A. Addazi et al., Eur. Phys. J. C **75** (2015) 400  [arXiv:1507.04317 [hep-ex]]

[37] D. Smith and N. Weiner, Phys. Rev. D **64** (2002) 043502  [arXiv:hep-ph/0101138]

[38] S. Chang, G.D. Kribs, D. Tucker-Smith, and N. Weiner, Phys. Rev. D **79** (2009) 043513 [arXiv:0807.2250 [hep-ph]]

[39] P. Finkbeiner, T. Lin, and N. Weiner, Phys. Rev. D **80** (2009) 115008 [arXiv:0906.0002 [astro-ph.CO]]

[40] R. Bernabei *et al.*, Mod. Phys. Lett. A **23** (2008) 2125 [arXiv:0802.4336 [astro-ph]] and references quoted therein

[41] R. Bernabei, presentation at the LNGS Scientific Committee, March 2018

[42] R. Bernabei *et al.,* Universe **4** no.11 (2018) 116, Nucl. Phys. Atom. Energy **19** no.4 (2018) [arXiv:1805.10486 [hep-ex]]

[43] J. Amaré *et al.* (ANAIS), arXiv:1903.03973 [astro-ph.IM]

[44] G. Adhikari *et al.* (COSINE), Eur. Phys. J. C **78** (2018) 107 [arXiv:1710.05299 [physics.ins-det]]

[45] K.-I. Fushimi *et al.* (PICO-LON),  arXiv:1605.04999 [astro-ph.IM]

[46] M. Antonello *et al.* (SABRE), Astrop. Phys**. 106** (2019) 1 [arXiv:1806.09340 [phys.ins-det]]

[47] F. Kahlhoefer et al. (COSINUS), JCAP **1805** no.5 (2018) 074 [arXiv:1802.10175 [hep-ph]]

[48] R. Bernabei et al., Eur. Phys. J. C **72** (2012) 2064 [arXiv:1202.4179 [astro-ph.GA]]

[49] R. Bernabei et al., Eur. Phys. J. C **74** (2014) 3196 [arXiv:1409.3516 [hep-ph]]

[50] A. H. Abdelhameed *et al.* (CRESST-III Collaboration), arXiv:1904.00498 [astro-ph.CO]

[51] E. Aprile *et al.* (XENON Collaboration), arXiv:1902.03234 [astro-ph.CO] and arXiv:1902.11297 [physics.ins-det]

[52] D.S. Akerib *et al*. (LUX Collaboration), arXiv:1811.11241 [astro-ph.CO]

[53] S. Baum, K. Freese, and C. Kelso, Phys. Lett. B **789** (2019) 262 [arXiv :1804.01231 [astro-ph.CO]]

[54] S. Kang, S. Scopel, G. Tomar, and J-H Yoon, JCAP **1807** no.7 (2018) 016 [arXiv:1804.07528[hep-ph]]

[55] A.L. Fitzpatrick, W. Haxton, E. Katz, N. Lubbers, and Y. Xu, JCAP **1302** (2013) 004 [arXiv:1203.3542 [hep-ph]]

[56] N. Anand, A.L. Fitzpatrick, and W. Haxton, Phys. Rev. C 89 (2014) no.6 065501 [arXiv:1308.6288 [hep-ph]]

[57] J. Herrero-Garcia, A. Scaffidi, M. White, and A.G. Williams, Phys. Rev. D **98** (2018) 123007 [arXiv:1804.08437 [hep-ph]]

[58] S. Kang, S. Scopel, G. Tomar, J.-H. Yoon, and P. Gondolo , JCAP **1811** no.11 (2018) 040 [arXiv:1808.04112 [hep-ph]]

[59] Talk given by P. Belli at *Particle Physics with Neutrons at the ESS,* Nordita, Stockholm (Sweden), December 10-14, 2018

[60] R. Bernabei et al., *Improved model-dependent corollary analyses after the first six annual cycles of DAMA/LIBRA − phase 2* (to appear)

[61] R. Bernabei *et al.* (DAMA Coll.), private communication